\documentclass{moriond}

\usepackage{cite}

\def\be{\begin{equation}}
\def\ee{\end{equation}}
\def\bea{\begin{eqnarray}}
\def\eea{\end{eqnarray}}

\newcommand{\Photo}{\includegraphics[height=35mm]{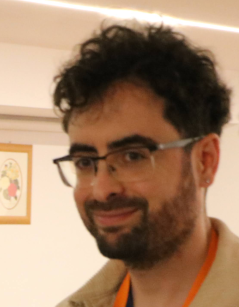}}

\usepackage{amsmath}
\usepackage{amsfonts}
\usepackage{amssymb}
\usepackage{mathtools}

\def\rd#1{{\color{red} #1}}

\begin{document}
\vspace*{4cm}
\title{Direct bounds on Left-Right gauge boson masses}

\author{ Sergio~Ferrando~Solera, Antonio~Pich, \underline{Luiz~Vale~Silva} }

\address{Departament de F\'{i}sica Te\`{o}rica, Instituto de F\'{i}sica Corpuscular, \\
Universitat de Val\`encia -- Consejo Superior de Investigaciones Cient\'{i}ficas, \\
Parc Cient\'{i}fic, Catedr\'{a}tico Jos\'{e} Beltr\'{a}n 2, E-46980 Paterna, Valencia, Spain}

\maketitle\abstracts{
%
%
%
%
While the third run of the Large Hadron Collider (LHC) is ongoing, 
the underlying theory 
that extends the Standard Model remains so far unknown.
Left-Right Models (LRMs) introduce a new gauge sector, and can restore parity symmetry at high enough energies.
If LRMs are indeed realized in nature,
the mediators of the new weak force can be searched for in colliders via their direct production.
We recast existing experimental limits
from the LHC Run~2
and derive generic bounds on the masses of the heavy LRM gauge bosons. 
As a novelty, we discuss the 
dependence of the $W_R$ and $Z_R$ total decay width on the LRM scalar content,
obtaining model-independent bounds within the specific realizations of the LRM scalar sectors analysed here. These bounds avoid the need to detail the spectrum of the scalar sector, and apply in the general case where no discrete symmetry is enforced.
Moreover, we emphasize
the impact on the $W_R$ production at LHC of general textures of the right-handed quark mixing matrix without manifest left-right symmetry.
We find that the $W_R$ and $Z_R$ masses are constrained to lie above $2$~TeV and $4$~TeV, respectively.
}

\section{Introduction}

The starting point is the enlarged gauge group $G_{LRM}$,
which is assumed to be spontaneously broken down to the Standard Model (SM) gauge group $G_{SM}$ at an energy scale $ |v_R| \gg v_{\rm EW} $, where $v_{\rm EW}$ sets the energy scale of electroweak (EW) Spontaneous Symmetry Breaking (SSB):

\vspace{1.5mm}
\begin{center}
\begin{tabular}{cc}
$ G_{LRM} = \underbrace{SU(3)_{\rm QCD}}_{g_s} \times \underbrace{SU(2)_L}_{g_L} \times \underbrace{SU(2)_R}_{g_R} \times \underbrace{U(1)_{X}}_{g_X} $ & \\
\multicolumn{2}{c}{LR-SSB @ scale $ v_{R} $, $\gamma$: mixing $W^3_R$ and $W_X$} \\
$ \downarrow $ & \\
$ G_{SM} = SU(3)_{\rm QCD} \times SU(2)_L \times U(1)_Y $ & \\
\multicolumn{2}{c}{EW-SSB @ scale $ v_{\rm EW} $, $\theta_W$: mixing $W^3_L$ and $W_Y$} \\ 
$ \downarrow $ & \\
$ SU(3)_{\rm QCD} \times U(1)_{\rm EM} $ & \\
\end{tabular}
\end{center}
\vspace{1.5mm}

\noindent
This category of models can restore a discrete symmetry among left- and right-handed (LH and RH, respectively) degrees of freedom, which in the SM is explicitly broken by EW interactions.
In principle, the energy scale of restoration of a discrete (e.g., parity) symmetry can be pushed to energy scales much above $|v_R|$, a possibility that will be considered hereafter \cite{Langacker:1989xa}. 
The quantum number of the Abelian factor $U(1)_{X}$ is identified with baryon minus lepton numbers $B-L$.
There is a mixing angle $\gamma$ analogous to the weak angle $\theta_W$ of the SM, and
there are new massive gauge bosons, called $W_R$ and $Z_R$, of masses proportional to the LR energy scale.
Working with perturbatively small couplings implies that the mixing angle $\gamma$ falls into a certain interval: considering that gauge couplings can value at most $\sim 1$, we have that \(70^{\circ}\gtrsim\gamma\gtrsim20^{\circ}\).
At low enough energies, corrections to the SM picture are suppressed by the ratio of the two energy scales, i.e., $ v_{\rm EW} / |v_R| $.

\begin{figure}[t]
    \centering
    \includegraphics[scale=0.26]{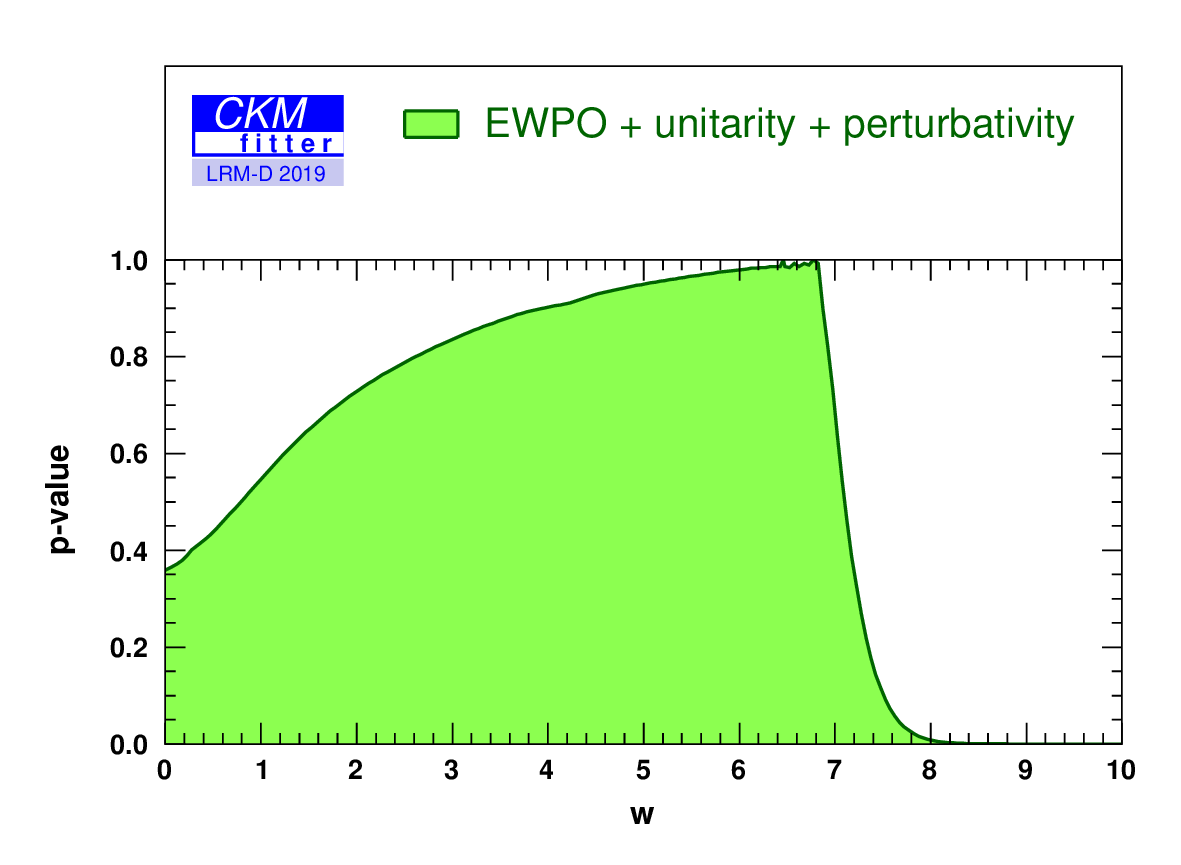}
    \includegraphics[scale=0.26]{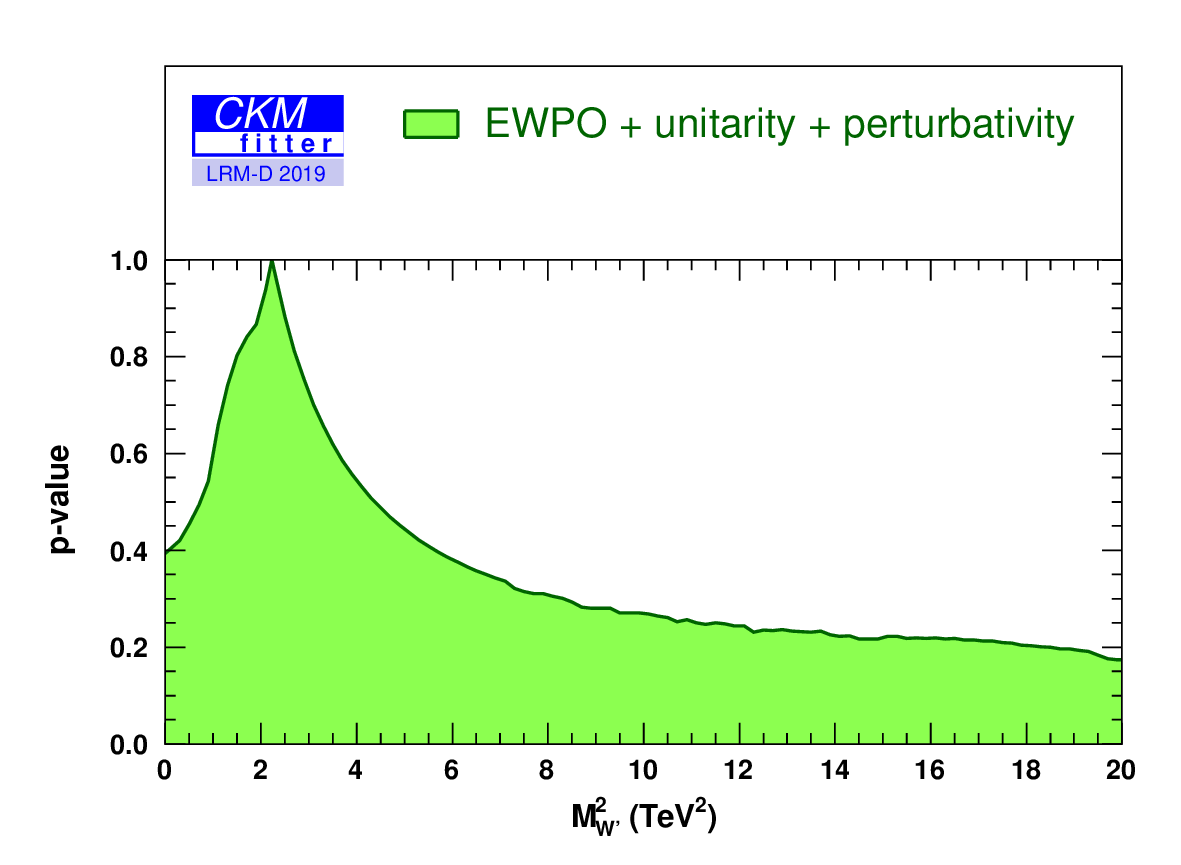}
    \includegraphics[scale=0.26]{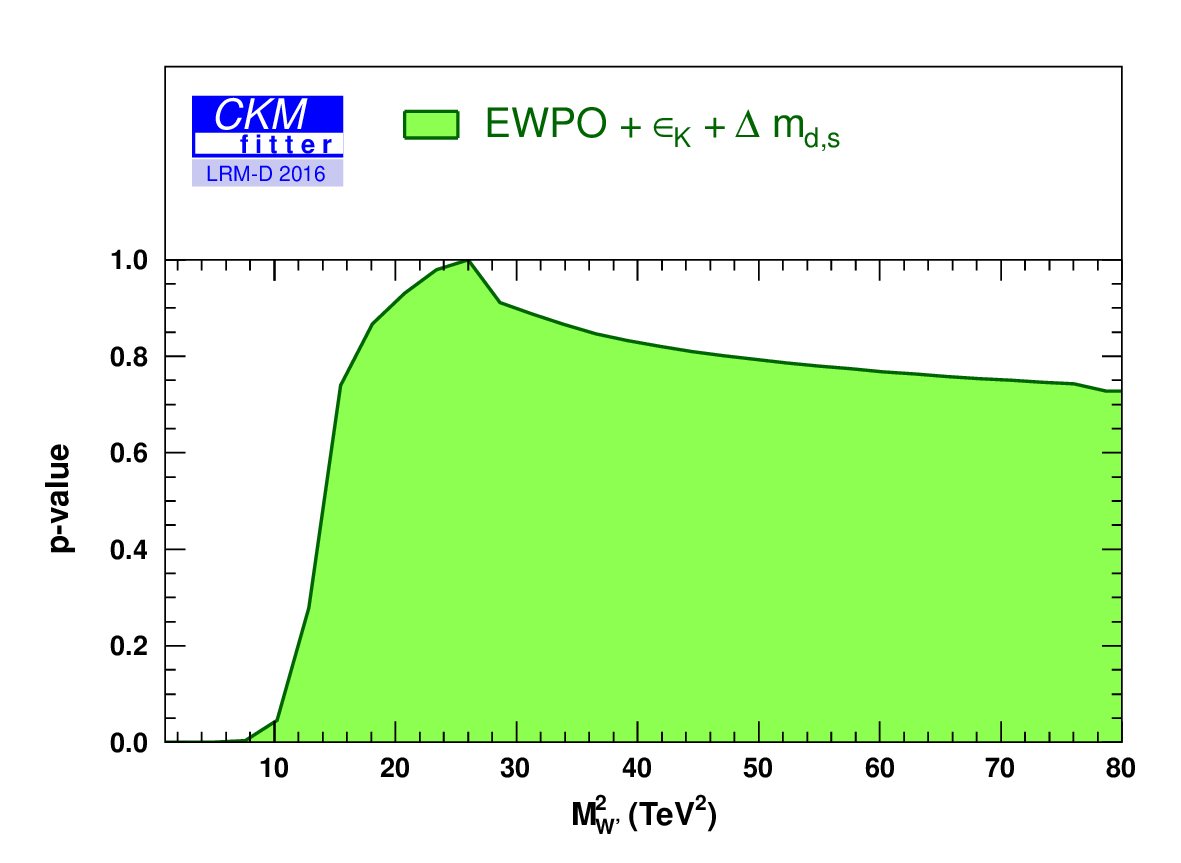}
    \caption{Bounds in the doublet scenario. (Left) Bound on the EW parameter $ w \propto |v_L|/v_{\rm EW} $. (Center) Bound on the $ W' \equiv W_R $ mass squared from EWPOs, and unitarity and perturbativity requirements. (Right) Bound on the $ W' \equiv W_R $ mass squared from EWPOs, and meson mixing in the manifest case.}
    \label{fig:EWPOs_flavour}
\end{figure}

The LR-SSB can be implemented in multiple ways, the most studied cases consist of a triplet or a doublet representation under $SU(2)_R$.
In the doublet scenario, a doublet representation under $SU(2)_L$ of vacuum expectation value (VEV) $v_L$ can implement the EW-SSB, as a global fit to EW precision observables (EWPOs) demonstrates \cite{Bernard:2020cyi}; see the left panel of Fig.~\ref{fig:EWPOs_flavour}, extracted from Ref.~\cite{Bernard:2020cyi}.
The possibility of a triplet representation under $SU(2)_L$ triggering EW-SSB is not allowed in the triplet scenario, since it would spoil the relation between the SM-like $W$ and $Z$ masses.
Additional scalar representations can be considered for multiple reasons.
For instance,
a bi-doublet (that is, a doublet under both $SU(2)_L$ and $SU(2)_R$) can be introduced for the sake of triggering EW-SSB in the triplet scenario.

Fermions in LRMs come in three copies of the following representations: 

\vspace{1.5mm}
\begin{center}
\begin{tabular}{cccc}
	& \textbf{Left} && \textbf{Right} \\
	$ SU(2)_L $ & $ \mathbf{2} $ && $ \mathbf{1} $ \\
	$ SU(2)_R $ & $ \mathbf{1} $ && $ \mathbf{2} $ \\
	$ \begin{matrix} {\rm quarks:} \\ B=1/3 \\ \end{matrix} $ & $ \begin{pmatrix} U_{L} \\ D_{L} \\ \end{pmatrix} $ && $ \begin{pmatrix} U_{R} \\ D_{R} \\ \end{pmatrix} $ \\
	$ \begin{matrix} {\rm leptons:} \\ L=-1 \\ \end{matrix} $ & $ \begin{pmatrix} \nu_{L} \\ \ell_{L} \\ \end{pmatrix} $ && $ \begin{pmatrix} \nu_{R} \\ \ell_{R} \\ \end{pmatrix} $ \\
\end{tabular}
\end{center}
\vspace{1.5mm}

\noindent
To generate fermion masses, one usually introduces a bi-doublet scalar.
Alternatively,
one can consider dimension-5 effective operators built from doublets to give masses to all fermions;
we will later refer to this scenario as the ``effective'' case, which could for instance be realized via the introduction of vector-like fermions, see Ref.~\cite{Babu:2018vrl} for a recent discussion. 
In contrast, the ``doublet'' (``triplet'') scenario discussed later, other than one doublet (triplet) under $SU(2)_R$ and one doublet (triplet) under $SU(2)_L$, also has one bi-doublet.
Note that
RH neutrinos are introduced in LRMs, in such a way that one achieves an anomalous-free theory.
In the doublet scenario, only Dirac mass terms are possible, while in the triplet scenario Majorana mass terms are also allowed: one implements in this way a see-saw mechanism in which RH neutrinos have masses proportional to $v_R$.

There are new mixing angles among the fermions; one has in particular a RH counterpart $V_R^{\rm CKM}$ to the SM-like CKM matrix $V_L^{\rm CKM}$.
When a discrete symmetry is enforced, the two mixing matrices in the quark sector are deeply connected.
For instance, we define the two following cases (see, e.g., Ref.~\cite{Maiezza:2010ic,Senjanovic:2014pva}):

\begin{eqnarray}
    & \mathcal{P} \text{ : manifest} \; \to & V_R^{\rm CKM} \simeq S_u V_L^{\rm CKM} S_d \,, \nonumber\\
    & \mathcal{C} \text{ : pseudo-manifest} \; \to & V_R^{\rm CKM} = K_u ( V_L^{\rm CKM} )^\ast K_d \,, \nonumber
\end{eqnarray}
where $ S_q $ ($ K_q $) are diagonal matrices carrying signs (complex phases; one is fixed by the others).

Before moving to collider physics, let us briefly comment on EWPOs and quark flavour physics.
EWPOs can be used to probe the properties of LRMs, however
we do not extract very constraining information on the model parameters; see the mid panel of Fig.~\ref{fig:EWPOs_flavour}, extracted from Ref.~\cite{Bernard:2020cyi}.
Moreover, an additional difficulty arises when computing LRM contributions beyond the tree level in order to get sensitivity to the scalar spectrum,
since a large number of parameters from the scalar potential intervene \cite{Bernard:2020cyi}.
In the case of quark flavour physics, let us first mention that there are no flavour-changing neutral currents (FCNCs) mediated by the $Z_R$ (at least in the models considered hereafter), while the scalar sector typically introduces FCNCs and is thus strongly constrained.
There is a rich phenomenology in LRMs emerging from RH charged currents, that can be probed for instance with the use of meson mixing \cite{Bernard:2015boz,Bertolini:2014sua}; see the right panel of Fig.~\ref{fig:EWPOs_flavour}, extracted from Ref.~\cite{ValeSilva:2016dcg}.
However, bounds strongly depend on the structure of the RH quark mixing matrix.
As a final comment, the contribution of the $W_R$ to the partial decay width of the SM-like Higgs to two photons does not set strong limits on the model \cite{Bandyopadhyay:2019jzq}.

\section{Collider bounds}

Moving now to collider data, there are many available searches for $W'$ and $Z'$.
To name a few, of more importance in the following: charged lepton pair \cite{ATLAS:2019erb,CMS:2021ctt}, di-jet \cite{ATLAS:2019fgd,CMS:2019gwf}, and charged lepton-neutrino pair \cite{ATLAS:2019lsy,CMS:2022krd} final states.
Many of these searches apply directly to categories of models other than LRMs, or to rather specific realizations of LRMs.
Here, we would like to consider LRMs in more generality, and discuss values for the gauge couplings and mixing matrix beyond the LR symmetric case.
We give more focus to fermionic decay modes;
different realizations of the scalar sector manifest then in the total decay widths of the $W' \to W_R$ and $Z' \to Z_R$, and also in the neutrino sector (i.e., whether RH neutrinos can possibly be too heavy, or else kinematically accessible).

Let us first discuss a necessary input, which is the total decay width $\Gamma$.
When computing $\Gamma$, one usually considers only the fermionic contribution.
This may be so because the non-fermionic decay width depends strongly on the particular realization of the scalar sector.
A simple expression for the upper bound on the total decay width can be obtained in the following way:
one considers the limit in which the whole set of scalars is kinematically accessible,
then discards the mass-dependent kinematical factor (which is bounded by a constant), and finally exploits the equivalence theorem. 
The resulting upper bound $\Gamma^{\rm upper} / M$
on $ \Gamma / M $ depends uniquely on the mixing angle $\gamma$.
Considering the value of $\Gamma$ that saturates the upper bound (namely, $\Gamma \to \Gamma^{\rm upper}$) leads to the smallest BRs, and thus to the most conservative bounds on the gauge boson masses.
The total decay width of the $Z_R$ in all three scenarios (doublet, triplet and the effective case) remains below 10\% of $M_{Z_R}$ for the interval of $\gamma$ allowed by the perturbative requirements considered above, while $\Gamma_{W_R}^{\rm upper}/M_{W_R}$ decreases with $g_R^2$ and is as small as 1\% for the smallest allowed value of $g_R \simeq 0.4$.
Luckily, we can exploit the fact that experimental searches consider different decay widths of the $Z_R$ and $W_R$.\footnote{
Compared to the purely fermionic decay width, $\Gamma_{W_R}^{\rm upper}$ is larger by 12.5\% and 25\% in the doublet and triplet cases (when RH neutrinos are included), respectively. In the case of $Z_R$, $\Gamma_{Z_R}^{\rm upper}$ is larger than the purely fermionic decay width by as much as a factor of $2$ in the triplet case for large values of the mixing angle $\gamma$ (again, when RH neutrinos are included).}
The recasting of existing experimental bounds corresponds then to a rescaling of the production and decay ratios for a given mass of the heavy gauge boson.

The bound on the mass of the $Z_R$ is dominated by searches of resonances going to a pair of charged leptons.
The cross section at the center-of-mass energy squared $s$ is given by

\begin{equation}
    \label{ecu.3.3}
    \sigma\left(pp\rightarrow Z_RX\rightarrow f\bar{f}X\right)\approx\frac{\pi}{6s}\sum_qc_q^{f}\,\omega_q\!\left(s,M_{Z_R}^2\right) \,,
\end{equation}

\noindent
where $\omega_q$ ($q = u, d, s, c, b, t$) is a function of the Parton Distributions Functions (PDFs), and the effective coefficients $c_q^{f}$ carry the important information about the production and decay strengths; see, e.g., Ref.~\cite{Accomando:2010fz}.
They depend as well on the value of $\gamma$.
The point that minimizes the cross section is not at the LR symmetric limit ($g_R = g_L$), such that by moving away from this limit the bound on the $Z_R$ mass is slightly relaxed.

A crucial input in the production and possibly in the decay of $W_R$ is the RH quark mixing matrix,
and as indicated by the expression of the cross section different production mechanisms accompany different elements of the RH quark mixing matrix:

\begin{equation}
    \label{ecu.3.13}
    \sigma\left(pp\rightarrow W_R^\pm X \right)\approx\frac{\pi}{6s} g_R^2 \sum_{ij}|\!\left(V_R^{\mathrm{CKM}}\right)_{ij}\!|^2\,\omega_{ij}\!\left(M^2_{W_R}/s,\,M_{W_R}\right) \,,
\end{equation}

\noindent
where $\omega_{ij}$ ($i = u, c, t$ and $j = d, s, b$) are functions of the PDFs, being displayed in Ref.~\cite{Bernard:2020cyi}.
In particular the $W_R$ production is larger when $ (V_R^{\rm CKM})_{u d} = 1 $.
However, if the texture of the mixing matrix is such that couplings to the lightest flavours are avoided, its production gets suppressed,
relaxing bounds on its mass by as much as $\sim 1$~TeV.
See also Ref.~\cite{Frank:2018ifw}.
In principle, considering associated production of $W_R$ with different heavy flavours could test the structure of the RH quark mixing matrix in colliders, if
one reconstructs the associated quark to limit the background from the SM.
However, some theoretical works indicate a limited sensitivity to gauge boson masses in the TeV range; see, e.g., Ref.~\cite{Frank:2010cj}.

Regarding direct searches based on the leptonic decay mode of the $W_R$,
many such searches consider heavy RH neutrinos of varying masses, and a particular structure for the mixing matrix in the leptonic sector is invoked.
(The problem in LRMs is different from the one dealt with in Ref.~\cite{Abada:2022wvh}, since here the mass of the charged gauge boson is also unknown.)
However, light neutrinos are possible, at least in the doublet case, but also in the triplet and effective scenarios.\footnote{
Being light, one sums over the different species, and the dependence on the leptonic mixing matrix drops.}

\begin{figure}[t]
    \centering
    \includegraphics[scale=0.423]{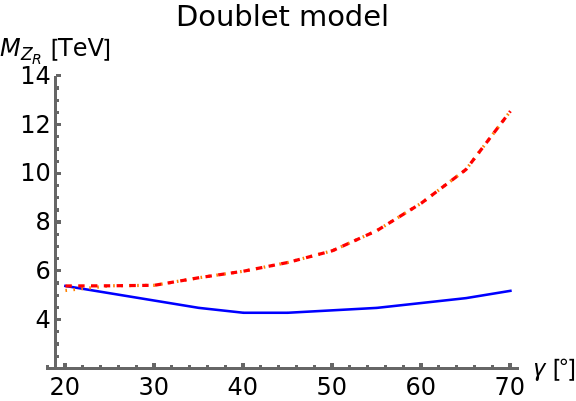} \hspace{10mm}
    \includegraphics[scale=0.423]{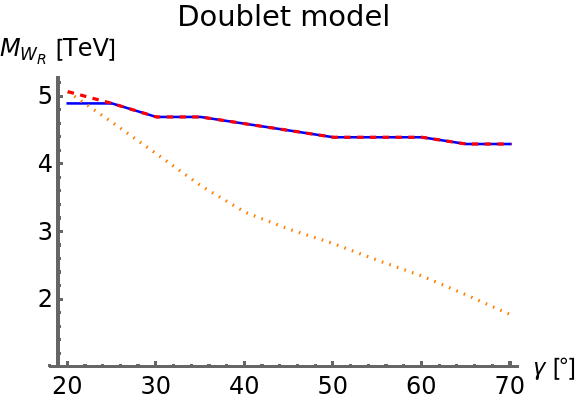} \\
    \vspace{5mm}
    \includegraphics[scale=0.33]{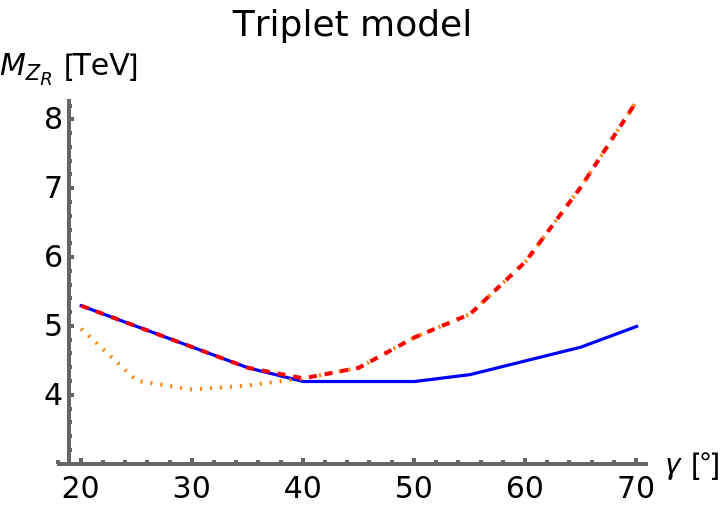} \hspace{10mm}
    \includegraphics[scale=0.33]{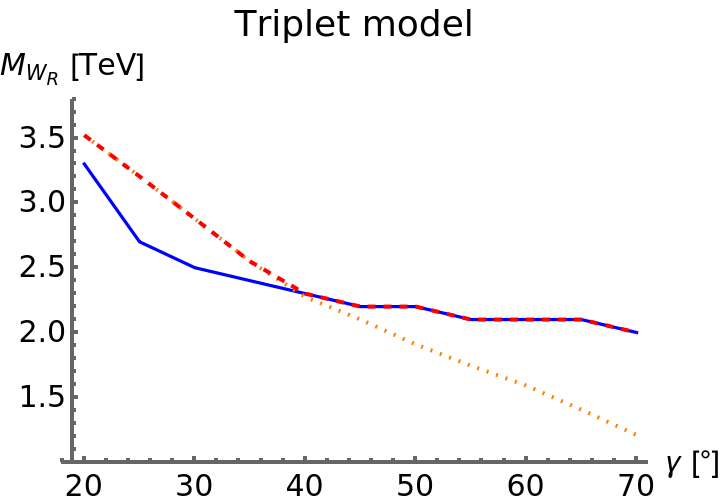}
    \caption{Bounds on $M_{Z_R}$ (left panels) and $M_{W_R}$ (right panels) in the doublet (upper panels) and triplet models (lower panels). The solid blue curves correspond to the bounds on $M_{Z_R}$ ($M_{W_R}$) derived from direct searches of the $Z_R$ (respectively, $W_R$) gauge boson; the dotted orange lines show the bounds on the $Z_R$ ($W_R$) mass derived from direct searches of the $W_R$ (respectively, $Z_R$) gauge boson, based on the relation between the two heavy gauge boson masses in the LRM realizations under discussion, see Eq.~\eqref{eq:relations_masses}. The strongest limits are displayed in dashed red. In the LR symmetric point $g_R = g_L$, the mixing angle $\gamma \approx 33^{\circ}$.}
    \label{fig:synergy_plot}
\end{figure}

Direct bounds as a function of $\gamma$ are displayed with solid blue curves in Fig.~\ref{fig:synergy_plot}.
The left and right panels show, respectively, the lower limits on $M_{Z_R}$ and $M_{W_R}$.
The $W_R$ figures correspond to an anti-diagonal RH quark mixing matrix that imposes less restrictive constraints.
In the triplet case, where Majorana neutrinos are possible, they are based on data from the $ W_R \to jj $ mode, which is independent of the neutrino sector.
In the doublet case, where neutrinos are purely Dirac fermions, the blue curve shows the more constraining bounds on $ M_{W_R} $ obtained
from the leptonic decay mode $ W_R \to \ell \bar{\nu}_R $ with light RH neutrinos.

Let us highlight the synergy between $W_R$ and $Z_R$ searches,
given that their masses are deeply connected; see also Ref.~\cite{Araz:2021dga}. Their relations are

\begin{equation}\label{eq:relations_masses}
    \text{doublet: } M_{W_R} = \cos \gamma \, M_{Z_R} \,, \quad \text{triplet: } \sqrt{2} \, M_{W_R} = \cos \gamma \, M_{Z_R} \,,
\end{equation}

\noindent
where the relative factor of $\sqrt{2}$ is due to the different LR-SSB mechanisms.
For instance, when the mixing angle $\gamma$ increases,
the indirect bound from $M_{W_R}$ tends to dominate the lower bound on the $Z_R$ mass;
this is illustrated in Fig.~\ref{fig:synergy_plot}
where a bound better by about a factor of $2$ is reached.
The opposite happens for small values of $\gamma$, in which case one can improve bounds on the $W_R$ mass based on bounds on $M_{Z_R}$.

The reader will find a more extensive discussion in Ref.~\cite{Solera:2023kwt}.
As discussed therein,
one does not see a big impact of the scalar realization from its manifestation in the total decay width.
However,
one sees in particular a significant impact from the structure of $V_R^{\rm CKM}$ as it has been discussed above; the decay mode (di-jet searches against di-lepton searches, which in the case of the $W_R$ depends on the RH neutrino sector parameters) also has an important impact.

\section{Conclusions}

We have discussed different realizations of LRMs, and we have considered that the LR discrete symmetry can be pushed to high energies, such that the gauge coupling $g_R$ and the mixing matrix $V_R^{\rm CKM}$ are free to vary.
Collider physics can provide powerful bounds independently of the specifics of the model.
We exploit many collider bounds, in particular for different total decay widths of the heavy gauge bosons.
Bounds can change substantially by considering different values of the gauge coupling and the texture of the unitary RH quark mixing matrix.
We have shown complementarity in the searches of $Z_R$ and $W_R$.
Even after relaxing bounds, one is still able to set lower bounds on their masses in the many TeV range; this is an impressive achievement by LHC, showing a strong competition with flavour bounds.


\section*{Acknowledgments}

We thank Prasanna K. Dhani, Andreas Hinzmann, Greg Landsberg, Jeongeun Lee, Emanuela Musumeci, Tamara Vazquez Schroeder, and Jos\'{e} Zurita for engaging in discussions and providing their comments.
This work has been supported by MCIN/AEI/10.13039/501100011033, grant PID2020-114473GB-I00; by Generalitat Valenciana, grant PROMETEO/2021/071 and by Ministerio de Universidades (Gobierno de Espa\~na), grant FPU20/04279.
This project has received funding from the European Union’s Horizon 2020 research and innovation programme under the Marie Sklodowska-Curie grant agreement No 101031558.
LVS is grateful for the hospitality of the CERN-TH group where part of this research was executed.

\section*{References}

\bibliography{mybib}{}

\begin{thebibliography}{10}

\bibitem{Langacker:1989xa}
Paul Langacker and S.~Uma Sankar.
\newblock {Bounds on the Mass of W(R) and the W(L)-W(R) Mixing Angle xi in General SU(2)-L x SU(2)-R x U(1) Models}.
\newblock {\em Phys. Rev. D}, 40:1569--1585, 1989.

\bibitem{Bernard:2020cyi}
V\'eronique Bernard, S\'ebastien Descotes-Genon, and Luiz Vale~Silva.
\newblock {Constraining the gauge and scalar sectors of the doublet left-right symmetric model}.
\newblock {\em JHEP}, 09:088, 2020.

\bibitem{Babu:2018vrl}
K.~S. Babu, Bhaskar Dutta, and Rabindra~N. Mohapatra.
\newblock {A theory of R(D$^{*}$, D) anomaly with right-handed currents}.
\newblock {\em JHEP}, 01:168, 2019.

\bibitem{Maiezza:2010ic}
Alessio Maiezza, Miha Nemevsek, Fabrizio Nesti, and Goran Senjanovic.
\newblock {Left-Right Symmetry at LHC}.
\newblock {\em Phys. Rev. D}, 82:055022, 2010.

\bibitem{Senjanovic:2014pva}
Goran Senjanovi\'c and Vladimir Tello.
\newblock {Right Handed Quark Mixing in Left-Right Symmetric Theory}.
\newblock {\em Phys. Rev. Lett.}, 114(7):071801, 2015.

\bibitem{Bernard:2015boz}
V\'eronique Bernard, S\'ebastien Descotes-Genon, and Luiz Vale~Silva.
\newblock {Short-distance QCD corrections to $ {K}^0{\overline{K}}^0 $ mixing at next-to-leading order in Left-Right models}.
\newblock {\em JHEP}, 08:128, 2016.

\bibitem{Bertolini:2014sua}
Stefano Bertolini, Alessio Maiezza, and Fabrizio Nesti.
\newblock {Present and Future K and B Meson Mixing Constraints on TeV Scale Left-Right Symmetry}.
\newblock {\em Phys. Rev. D}, 89(9):095028, 2014.

\bibitem{ValeSilva:2016dcg}
Luiz Vale~Silva.
\newblock {\em {Phenomenology of Left-Right Models in the quark sector}}.
\newblock PhD thesis, Saclay, 2016.

\bibitem{Bandyopadhyay:2019jzq}
Triparno Bandyopadhyay, Dipankar Das, Roman Pasechnik, and Johan Rathsman.
\newblock {Complementary bound on the $W^\prime$ mass from Higgs boson to diphoton decays}.
\newblock {\em Phys. Rev. D}, 99(11):115021, 2019.

\bibitem{ATLAS:2019erb}
Georges Aad et~al.
\newblock {Search for high-mass dilepton resonances using 139 fb$^{-1}$ of $pp$ collision data collected at $\sqrt{s}=$13 TeV with the ATLAS detector}.
\newblock {\em Phys. Lett. B}, 796:68--87, 2019.

\bibitem{CMS:2021ctt}
Albert~M Sirunyan et~al.
\newblock {Search for resonant and nonresonant new phenomena in high-mass dilepton final states at $ \sqrt{s} $ = 13 TeV}.
\newblock {\em JHEP}, 07:208, 2021.

\bibitem{ATLAS:2019fgd}
Georges Aad et~al.
\newblock {Search for new resonances in mass distributions of jet pairs using 139 fb$^{-1}$ of $pp$ collisions at $\sqrt{s}=13$ TeV with the ATLAS detector}.
\newblock {\em JHEP}, 03:145, 2020.

\bibitem{CMS:2019gwf}
Albert~M Sirunyan et~al.
\newblock {Search for high mass dijet resonances with a new background prediction method in proton-proton collisions at $\sqrt{s} =$ 13 TeV}.
\newblock {\em JHEP}, 05:033, 2020.

\bibitem{ATLAS:2019lsy}
Georges Aad et~al.
\newblock {Search for a heavy charged boson in events with a charged lepton and missing transverse momentum from $pp$ collisions at $\sqrt{s} = 13$ TeV with the ATLAS detector}.
\newblock {\em Phys. Rev. D}, 100(5):052013, 2019.

\bibitem{CMS:2022krd}
Armen Tumasyan et~al.
\newblock {Search for new physics in the lepton plus missing transverse momentum final state in proton-proton collisions at $\sqrt{s} =$ 13 TeV}.
\newblock {\em JHEP}, 07:067, 2022.

\bibitem{Accomando:2010fz}
Elena Accomando, Alexander Belyaev, Luca Fedeli, Stephen~F. King, and Claire Shepherd-Themistocleous.
\newblock {Z' physics with early LHC data}.
\newblock {\em Phys. Rev. D}, 83:075012, 2011.

\bibitem{Frank:2018ifw}
Mariana Frank, \"Ozer \"Ozdal, and Poulose Poulose.
\newblock {Relaxing LHC constraints on the $W_R$ mass}.
\newblock {\em Phys. Rev. D}, 99(3):035001, 2019.

\bibitem{Frank:2010cj}
Mariana Frank, Alper Hayreter, and Ismail Turan.
\newblock {Production and Decays of $W_R$ bosons at the LHC}.
\newblock {\em Phys. Rev. D}, 83:035001, 2011.

\bibitem{Abada:2022wvh}
Asmaa Abada, Pablo Escribano, Xabier Marcano, and Gioacchino Piazza.
\newblock {Collider searches for heavy neutral leptons: beyond simplified scenarios}.
\newblock {\em Eur. Phys. J. C}, 82(11):1030, 2022.

\bibitem{Araz:2021dga}
Jack~Y. Araz, Mariana Frank, Benjamin Fuks, Stefano Moretti, and \"Ozer \"Ozdal.
\newblock {Cross-fertilising extra gauge boson searches at the LHC}.
\newblock {\em JHEP}, 11:014, 2021.

\bibitem{Solera:2023kwt}
Sergio~Ferrando Solera, Antonio Pich, and Luiz Vale~Silva.
\newblock {Direct bounds on Left-Right gauge boson masses at LHC Run 2}.
\newblock {\em JHEP}, 02:027, 2024.

\end{thebibliography}

\end{document}